\documentclass[prb,twocolumn,superscriptaddress,floatfix,noshowpacs,10pt,longbibliography,nofootinbib]{revtex4-2}
\usepackage{graphicx,bm,times}
\usepackage{amsmath}
\usepackage{amsfonts}
\usepackage{amssymb}
\usepackage[version=4]{mhchem} 
\usepackage{lipsum}
\usepackage{bm}
\usepackage{color}
\usepackage{times}
\usepackage{xcolor}
\usepackage{soul}
\usepackage{verbatim}
\usepackage[normalem]{ulem}
\useunder{\uline}{\ul}{}

\usepackage[%
  colorlinks=true,
  urlcolor=blue,
  linkcolor=blue,
  citecolor=blue
  ]{hyperref}
\DeclareUnicodeCharacter{03B1}{\alph}

\newif\ifptitle
\newif\ifpnumber
\newcounter{para}

\ptitletrue  
\pnumbertrue  

\definecolor{mRed}{RGB}{230, 0, 50}
\colorlet{newtextColor}{mRed}
\newif\iftrackchanges

\trackchangestrue  

\begin{document}

\title{Angular Modulations in Magnetic Torque Induced by Phase Transitions \\ in the Triangular Supersolid $2H$-AgNiO$_2$}

\author{John~S.~Pearce}
\email[corresponding author:]{john.pearce@physics.ox.ac.uk}
\affiliation{Clarendon Laboratory, Department of Physics,
	University of Oxford, Parks Road, Oxford OX1 3PU, UK}

\author{Zeyu~Ma}
\altaffiliation[Present address: ]{Harvard John A. Paulson School of Engineering and Applied Sciences, Oxford Street, Cambridge, MA 02138}
\affiliation{Clarendon Laboratory, Department of Physics,
	University of Oxford, Parks Road, Oxford OX1 3PU, UK}

\author{Alimamy Bangura}
\affiliation{National High Magnetic Field Laboratory,
Florida State University, Tallahassee, FL 32310, USA}

\author{Timo Sörgel}
\affiliation{Max-Planck-Institut fur Festkörperforschung, Heisenbergstr. 1,
70569 Stuttgart, Germany}

\author{Martin Jansen}
\affiliation{Max-Planck-Institut fur Festkörperforschung, Heisenbergstr. 1,
70569 Stuttgart, Germany}

\author{Radu~Coldea}
\affiliation{Clarendon Laboratory, Department of Physics,
University of Oxford, Parks Road, Oxford OX1 3PU, UK}

\author{Amalia~I.~Coldea}
\email[corresponding author:]{amalia.coldea@physics.ox.ac.uk}
\affiliation{Clarendon Laboratory, Department of Physics,
University of Oxford, Parks Road, Oxford OX1 3PU, UK}

\date{\today}

\begin{abstract}

Easy-axis frustrated triangular lattice antiferromagnets provide an important playground for stabilising a wide range of exotic magnetic phases. The delafossite $2H$-AgNiO$_2$ offers a unique model system containing a lattice of localised ($S=1$) moments surrounded by a honeycomb of itinerant electrons.
Below $T_\text{N} =19$~K, the system exhibits collinear stripe antiferromagnetic order, whereas applied magnetic fields induce a cascade of transitions which has been proposed to contain a magnetic supersolid phase. 
In this study, we perform detailed angular and field-dependent torque measurements in static fields of up to 45~T on single crystal samples.
The angular dependence of the torque displays a sawtooth-shaped signal close to the first supersolid phase, while near $T_\text{N}$, an unexpected additional modulation emerges.
On the other hand, the field dependence of the torque indicates the presence of four field-induced anomalies which evolve in distinct ways as a function of field orientation.
To understand these complex behaviours and assess the role of local moments, we employ mean-field and Monte-Carlo calculations on a minimal spin model and construct comprehensive polar phase diagrams at different temperatures.
We find that the angular dependence of the torque can be understood in terms of crossing different phase boundaries by field rotation, and we identify qualitative similarities between the experimental and theoretical polar phase diagrams.
Our study highlights $2H$-AgNiO$_2$ as a rich anisotropic system that provides a robust example of how spin supersolid orders 
can manifest in angle-dependent magnetic torque.
\end{abstract}

\maketitle

\section{Introduction}
Despite its simplicity, the triangular lattice antiferromagnet still constitutes a significant portion of frustrated magnetism research, owing to the accessible platform it provides for investigating a vast range of exotic magnetic phenomena. This includes, but is not limited to, the resonance valence bond spin liquid \cite{Anderson1973,Yang2021,Ranjith2019} and the 120$^\circ$ Néel ordered phase \cite{Seabra2011_2, Capriotti1999}. Increasing the complexity of the nearest-neighbor triangular antiferromagnet model, via tuning second neighbor interactions or incorporating single-ion anisotropy, only serves to diversify the already rich phase diagram, opening avenues to collinear stripe orders and incommensurate spin helices \cite{Lecheminant1995,Jolicoeur1990,Wheeler2009}.

A \textit{supersolid} is a phase of matter that breaks both translational and $U(1)$ rotational symmetry \cite{Lifshitz1969}. In the original context of defects in solids, supersolids were predicted to arise from the Bose-Einstein condensation of solid defects with sufficiently large zero-point motion \cite{Lifshitz1969}, as was thought to be the case in $^4$He \cite{Boninsegni2012, Meisel1992, Kim2004}. A magnetic analogue of a supersolid, 
resulting instead in the breaking of lattice and spin-rotational symmetries, may also be stabilised in frustrated triangular magnets \cite{Melko2005, Wang2009, Seabra2010, Seabra2011}.
However, the required intimate tuning of anisotropic exchange interactions \cite{Melko2005, Seabra2010, Seabra2011} makes the magnetic supersolid difficult to realize experimentally, and few candidates have been identified, notably Na$_2$BaCo(PO$_4$)$_2$ \cite{Xiang2024} and recently A$_2$Co(SeO$_3$)$_2$ (A = K, Rb) \cite{Cui2025, Xu2025, Shi2025, Chen2026}.

\begin{figure*}[hbtp]
	\centering
\includegraphics[width=\linewidth,clip=true]{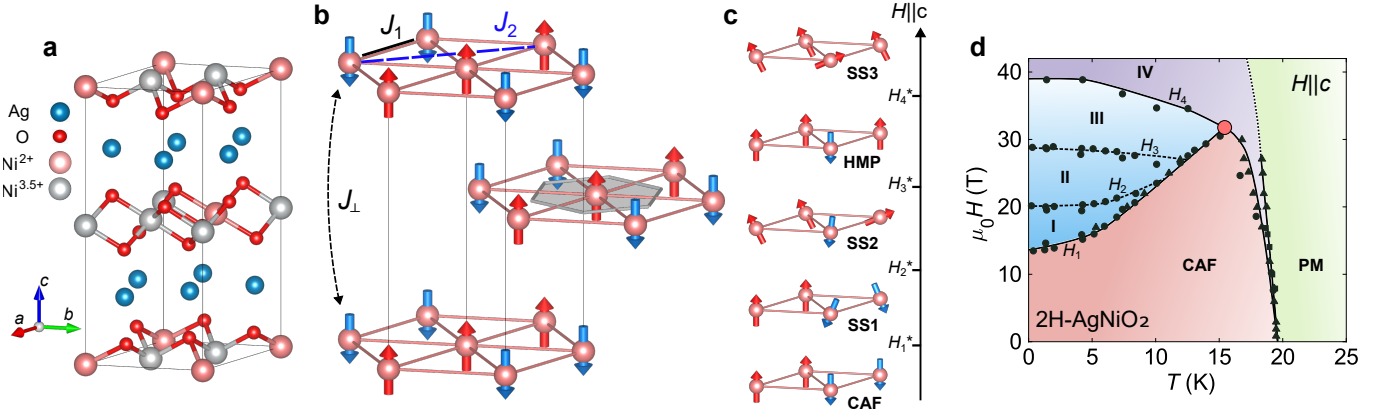}
\vspace{-0.2cm}
\caption{{\bf Magnetic ordering in $2H$-AgNiO$_2$.}
{\bf (a)} Crystallographic unit cell (space group $P6_322$).
{\bf (b)} Collinear antiferromagnetic stripe ordering of Ni$^{2+}$ ions (pink) on triangular layers. The shaded gray hexagon depicts a piece of itinerant Ni$^{3.5+}$ honeycomb surrounding a Ni$^{2+}$ spin. The full Ni$^{3.5+}$ honeycomb has been omitted for clarity. $J_{1,2,\perp}$ are Heisenberg exchange interactions, corresponding to the first nearest-neighbour, second nearest-neighbour, and interlayer interactions, respectively.
{\bf (c)} Semi-classical cartoons depicting the collinear antiferromagnet (CAF), first supersolid (SS1), 2:1:1 canted supersolid (SS2), half-magnetisation plateau (HMP), and 3:1 canted supersolid (SS3) structures, predicted by mean field theory at absolute zero, after Refs.~\cite{Seabra2010,Seabra2011}. The cartoons are ordered from bottom to top with respect to increasing magnetic field applied along the easy $c$ axis.
{\bf (d)} Experimental $H$-$T$ phase diagram of $2H$-AgNiO$_2$ (with $H||c$) after Ref.~\cite{Coldea2014}. The solid red circle highlights a tri-critical point. The paramagnetic and collinear antiferromagnetic regions are labelled with PM and CAF, respectively. The regions I-IV relate to magnetic phase transitions discussed in the main text and in Ref.~\cite{Coldea2014}.
}
\label{Fig1}
\end{figure*}

Much theoretical work has been done regarding supersolidity in the case of interacting hard-core bosons on a triangular lattice, where the problem has been directly mapped to a geometrically frustrated spin-1/2 \textit{XXZ} model \cite{Wang2009, Melko2005}. Quantum Monte Carlo simulations show that the supersolid order can be understood as arising via an order-by-disorder mechanism \cite{Melko2005,Wang2009}.
Alternatively, one can consider a semi-classical easy-axis $J_1$-$J_2$ model 
which reveals a rich phase diagram with multiple supersolid orders that depend sensitively on the single-ion anisotropy \cite{Seabra2010,Seabra2011}.

Although most studies of triangular lattice antiferromagnets are based on insulators, hexagonal $2H$-AgNiO$_2$ stands out as an uncommon \textit{metallic} example \cite{Sorgel2005, Coldea2014}.
A charge ordering transition below $T_\text{CO}$ = 365~K
splits the Ni sites into two interwoven sublattices (Fig.~\ref{Fig1}(a)), the first being a layered Ni$^{2+}$ ($S=1$) triangular antiferromagnet (Fig.~\ref{Fig1}(b)) and the second an itinerant honeycomb network of Ni$^{3.5+}$ ions \cite{Wawrzynska2008}. Band structure calculations show that such a structure can be thought of, almost literally, as a triangular antiferromagnetic insulator superimposed on a honeycomb metal \cite{Wawrzynska2008}. Neutron diffraction reveals a stripe collinear ground state which can be explained by a $J_1$-$J_2$ Heisenberg model \cite{Wawrzynska2008, Wheeler2009}. In addition, inelastic neutron scattering data suggests the existence of a moderate easy-$c$-axis anisotropy \cite{Wheeler2009,Coldea2014}.

Experimentally, $2H$-AgNiO$_2$ has a rich field-temperature phase diagram (Fig.~\ref{Fig1}(d)) \cite{Coldea2009, Coldea2014} where a zero-field paramagnetic (PM) to collinear antiferromagnetic (CAF) transition occurs at $T_\text{N}\approx19$~K. Applied magnetic field along the easy axis inside the CAF phase induces numerous anomalies in both magnetic torque and electronic transport \cite{Coldea2014}.
Among these, the first field-induced phase (I)
was associated with a supersolid where, semi-classically, 
down spins of the CAF order cant continuously away from the easy axis \cite{Coldea2014, Seabra2010, Seabra2011}, in stark contrast to the spin-flop transition observed in typical antiferromagnets \cite{Manson2023, Bogdanov2007}.
On the other hand, the microscopic mechanism responsible for anomalies attached to phase transitions I--II and II--III has remained elusive \cite{Coldea2014}.

In this study, we perform an angular-dependent torque magnetometry study in static magnetic fields of up to 45~T and temperatures down to 400~mK, to directly assess the stabilisation of different magnetic phases in $2H$-AgNiO$_2$. 
To understand the anomalies observed in magnetic torque, we compare our results to mean-field calculations and classical Monte Carlo simulations on a minimal easy-axis triangular $J_1$-$J_2$ model at different temperatures.
We find that the angular dependence of the torque can be qualitatively explained, within the local spin model, by considering transitions across different phase boundaries induced by rotating the applied magnetic field.

\section{Experimental Details}
$2H$-AgNiO$_2$ single crystals were synthesised under high oxygen pressure \cite{Sorgel2005} and screened via single-crystal X-ray diffraction. The observed diffraction patterns were consistent with the established low-temperature crystal structure space group $P6_322$ (no. 182), as shown in Fig.~S1 in the Supplementary Material (SM) \cite{SM}. The samples chosen for torque experiments were small platelets no greater than approximately $50\times50\times5$~$\mu$m$^3$ in size. Torque measurements were performed on three high-quality samples (S1, S2 and S3) mounted onto SEIKO PRC400 and PRC120 piezocantilevers. The cantilevers were glued to a cryogenic rotator platform and were cooled down to 2~K using a Quantum Design PPMS for measurements up to 16~T, and down to 400~mK in a $^3$He cryostat at NHMFL Tallahassee, up to 45~T. The platforms were rotated about a single axis perpendicular to the direction of the applied field. The samples were positioned on the cantilevers in two distinct orientations such as to probe the angular dependence of the torque in two orthogonal crystallographic planes, $ac$ and $a^*c$, where $a$ and $c$ reflect the conventional axes of the hexagonal $P6_322$ space group. The torque was calibrated in absolute units using a constant of proportionality determined in previous experiments by measuring the torque from the weight of a $\sim$4.5~$\mu$g sample in zero field. Details of the calibration may be found in the supplementary material (SM Fig.~S8 \cite{SM}).

For mean-field calculations of the magnetic ground state in a given field, we equilibrate a $2\times2\times1$ magnetic supercell of classical spins using a steepest-descents method \cite{LaBonte1969,Exl2014} implemented in the software library, SpinW \cite{spinW}. For supporting simulations at finite temperatures, we implement a parallel tempering Monte Carlo method \cite{Seabra2011, Hukushima1996, Kanki2005}, details of which are available in the SM \cite{SM}.

 \begin{figure}[hbtp]
	\centering
\includegraphics[width=\linewidth, clip=true]{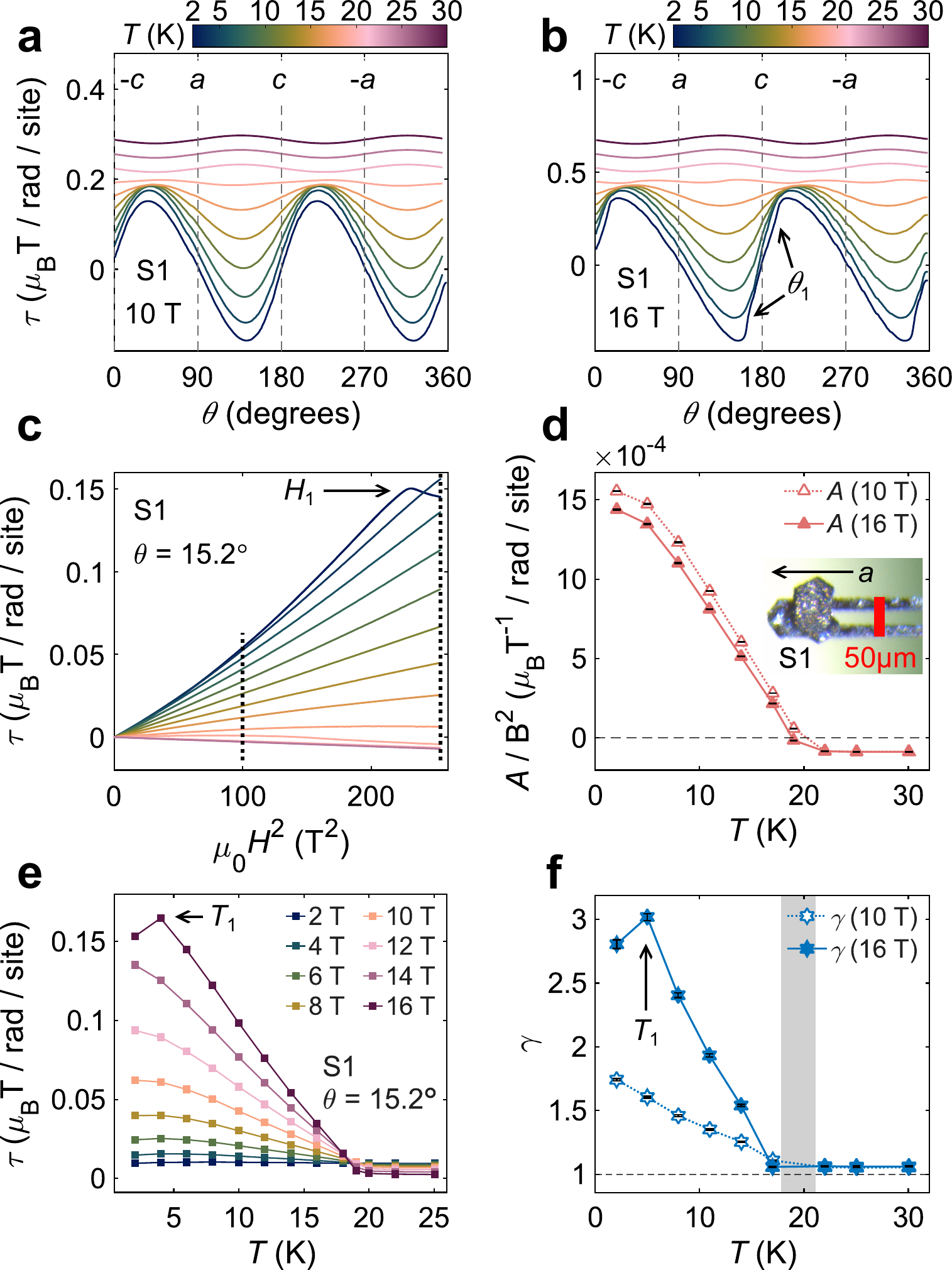}
\vspace{-0.2cm}
		\caption{
  {\bf Evolution of the angular dependence of the magnetic torque with temperature.}
  {\bf (a, b)} Angular dependence of the magnetic torque $\tau$ of sample S1 at different temperatures and in constant magnetic fields of 10~T and 16~T, respectively. $\theta$ is the angle of the field with respect to the crystallographic $c$ axis for field rotations within the $ac$ plane. Traces have been offset vertically for clarity. Dashed vertical lines indicate where the magnetic field is aligned with crystallographic axes. In panel (b), the arrows labelled $\theta_1$ point to kinks in $\tau$ observed near the easy $c$ axis at 2~K.
  {\bf (c, e)} Torque versus (c) field squared and (e) temperature at a constant $\theta=15^\circ$ away from the $c$ axis. In panel (c), the dotted black lines correspond to 10~T and 16~T, and the colour mapping is the same as in panels (a) and (b). The arrows labelled $H_1$ and $T_1$ highlight anomalies in $\tau$ which are related to crossing the CAF--I phase boundary by varying field and temperature, respectively.
  {\bf (d, f)} Fitted amplitude, $A$, and sawtooth parameter, $\gamma$, versus temperature in constant fields of 10~T and 16~T.  The gray region highlights where the torque is dominated by additional Fourier components, and hence cannot be fitted to a two-fold sawtooth model. The inset of (d) shows sample S1 mounted onto a piezocantilever with an arrow to show the direction of its $a$ axis. In panel (f), $T_1$ coincides with an anomaly in $\gamma$ in 16~T.
}
  \label{Fig2}
\end{figure}

\section{Results and Discussion}

\subsection{Angular dependence of the magnetic torque}
The angular dependence of the magnetic torque for sample S1 in constant field (10~T and 16~T) at different temperatures for rotation in the $ac$ plane perpendicular to the honeycomb layers is shown in 
Figs.~\ref{Fig2}(a) and (b). Above $T_\text{N}\approx$~19~K, the torque has a pure two-fold sinusoidal shape that decreases in amplitude as the temperature is lowered towards 19~K. Below 19~K, the sign of the torque is flipped, and the amplitude of the torque increases upon further cooling to base temperature. A similar behaviour is also observed in the angular dependence of the torque for sample S2, as shown in Fig.~S9 in the SM \cite{SM}.

To understand these changes in torque, we first need to clarify its sensitivity 
to magnetic anisotropy.
The magnetic torque
results from the misalignment between magnetisation $\mathbf{M}$ (per site) and applied magnetic field $\mathbf{H}$. Assuming that $\mathbf{H}$ and $\mathbf{M}$ both lie within the $ac$ plane (see Fig.~\ref{Fig2}a), the torque is $\tau=\mu_0|\mathbf{M}\times \mathbf{H}|\propto (\chi_{a}-\chi_c)H^2 \sin{2\theta}$, where $\theta$ is the angle subtended by $\mathbf{H}$ and the $c$ axis \cite{Coldea2014}.
The torque, as a derivative of the free energy,
$\tau=-\partial F/\partial\theta$, also confirms that a stable equilibrium position with $\partial\tau/\partial\theta<0$ 
is reached when magnetic field 
is aligned along the $c$ axis at high temperatures. This assignment is consistent with the easy axis anisotropy of $2H$-AgNiO$_2$ \cite{Wheeler2009, Wawrzynska2008}. 
Neutron diffraction showed that $2H$-AgNiO$_2$ enters a CAF ground state below $T_\text{N}\approx19$~K \cite{Wawrzynska2008}, at which point the susceptibility anisotropy between the crystallographic $a$ and $c$ axes, $(\chi_{a}-\chi_c)$, is expected to change sign \cite{Keffer1952}. This is confirmed by the sign change of the magnetic torque at $T_\text{N}$ apparent in Figs.~\ref{Fig2}(a), (b), (d), and (e).

The field dependence of the magnetic torque for a fixed  orientation $\theta=15^\circ$ away from the easy $c$ axis is shown in Fig.~\ref{Fig2}(c).
The torque exhibits a dominantly linear relationship with the squared magnetic field strength, $\tau\propto H^2$, 
 as expected for small $H$ inside the CAF and PM phases.
However, at 2~K, a stark change in slope is observed in the field dependence of the torque at the CAF--I phase boundary, $H_1\approx14$~T, as shown in Fig.~\ref{Fig2}(c) (top arrow). Meanwhile, a kink also appears in the temperature dependence of the torque in 16~T at $T_1\approx$4~K, as shown in Fig.~\ref{Fig2}(e).
These findings are 
entirely consistent with the proposed CAF--supersolid transition observed in previous studies \cite{Coldea2009, Coldea2014}.

{\bf Sawtooth behaviour of the torque}
Figs.~\ref{Fig2}(a) and (b) show that the sinusoidal waveform of the torque tilts into a sawtooth shape at lower temperatures, reflecting the growth of magnetic anisotropy \cite{Riedl2019, Ma2024}. Sawtooth behaviour in magnetic torque has been observed before, for example in Kitaev honeycomb magnets such as $\alpha$-RuCl$_3$ and $\alpha$-RuI$_3$ \cite{Froude2024, Ma2024}, and may be understood as resulting from the strong influence of anisotropic terms in the spin Hamiltonian at low temperatures \cite{Riedl2019}. 
To account for this behaviour, we employ
a phenomenological description of torque of the form \cite{Ma2024,Pearce2024}:
\begin{equation}
\label{eqn:sawtooth}
\tau(\theta) = A \left(\frac{\gamma+1}{2}\right) \frac{\sin{2\theta}}{\sqrt{\cos^2{\theta} + {\gamma^2}\sin^2{\theta}}}
\end{equation}
where $A$ is the amplitude and $\gamma$ parametrises the sawtooth shape ($\gamma=1$ for a pure sinusoid).
Figs.~\ref{Fig2}(d) and (f) show the temperature dependence of $A$ and $\gamma$, where $\gamma=1$ above $T_\text{N}=19$~K. Below $T_\text{N}$, $\gamma$ increases linearly with decreasing temperature, although a clear kink is observed in $\gamma$ at $T_1\approx4$~K in a field strength of 16~T. Although Eqn.~\ref{eqn:sawtooth} accurately captures the torque angular dependence in 10~T (Fig.~\ref{Fig2}(a)), it does not fully account for it in 16~T, where the fit residual reveals significant deviations from the model close to the easy $c$ axis, as evident in Fig.~S11 in the SM \cite{SM}. In 16~T, these deviations are caused by subtle anomalies that manifest close to the easy $c$ axis below 4~K (labelled $\theta_1$ in Fig.~\ref{Fig2}(b)). The anomalies are characterised by jumps in the first derivative, $\partial\tau/\partial\theta$, as shown in Fig.~S12 in the SM \cite{SM}.

\begin{figure}[hbtp]
	\centering
\includegraphics[width=\linewidth, clip=true]{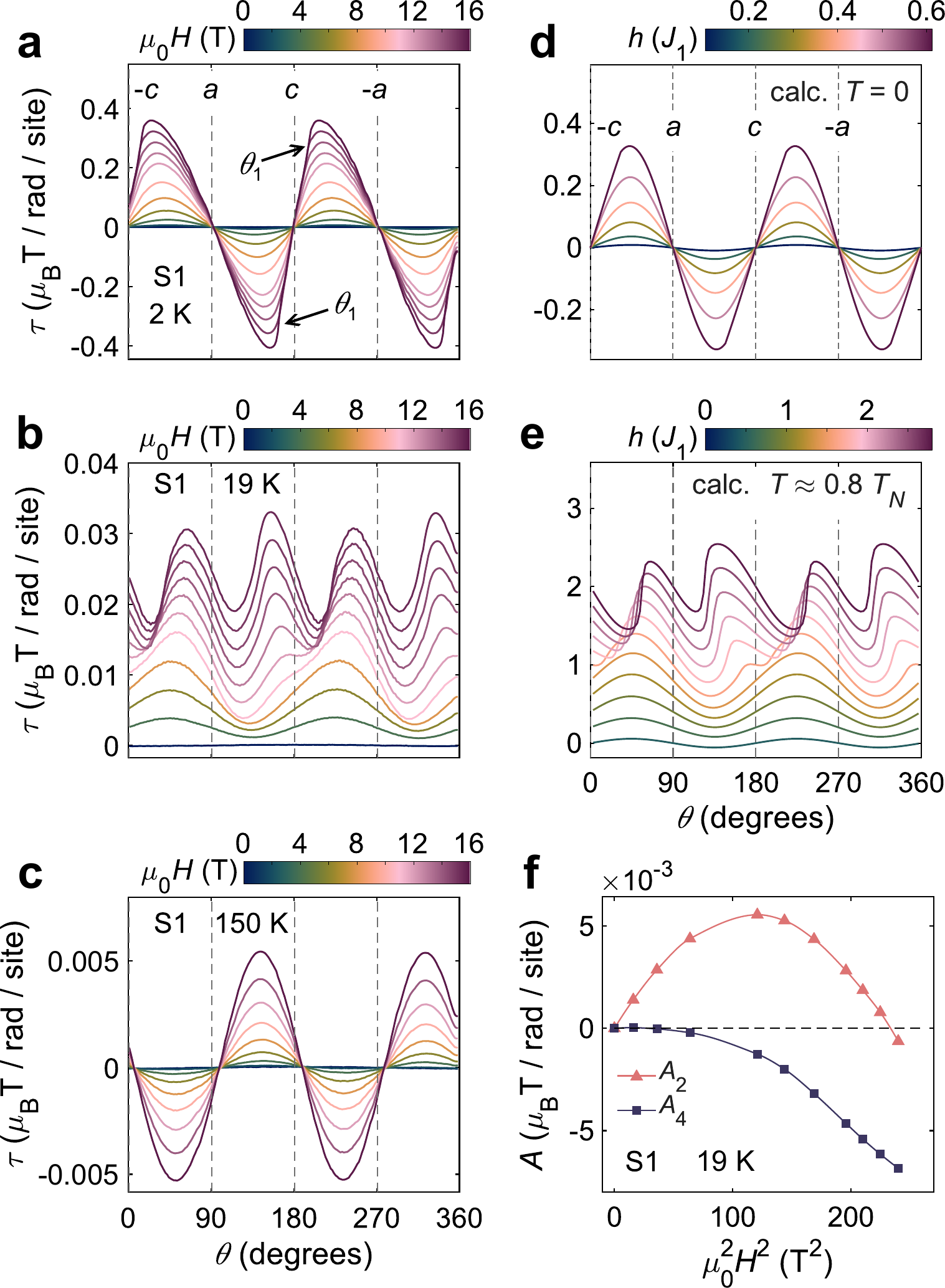}
\vspace{-0.2cm}
		\caption{
  {\bf Evolution of the angular dependence of magnetic torque with magnetic field.}
  {\bf (a-c)} Angular dependence of the magnetic torque for sample S1 in different magnetic field strengths up to 16~T and at constant temperatures of 2, 19, and 150~K, respectively. $\theta$ denotes the field rotation angle with respect to the crystallographic $-c$ axis in the $ac$ plane. Traces have been offset vertically for clarity in panel (b). In (a), the arrows labelled $\theta_1$ point to kinks that appear near the easy $c$ axis above 14~T.
  {\bf (d, e)} Calculated angular dependence of the magnetic torque at $T=0$ and $0.8~T_N$, respectively, using mean field theory and Monte Carlo. Traces are offset vertically for clarity in panel (e). {\bf (f)} The measured two-fold ($A_2$) and four-fold ($A_4$) Fourier amplitudes versus squared magnetic field strength at 19~K.
  }
  \label{Fig3}
\end{figure}

\begin{figure*}[hbtp]
	\centering
\includegraphics[width=\linewidth, clip=true]{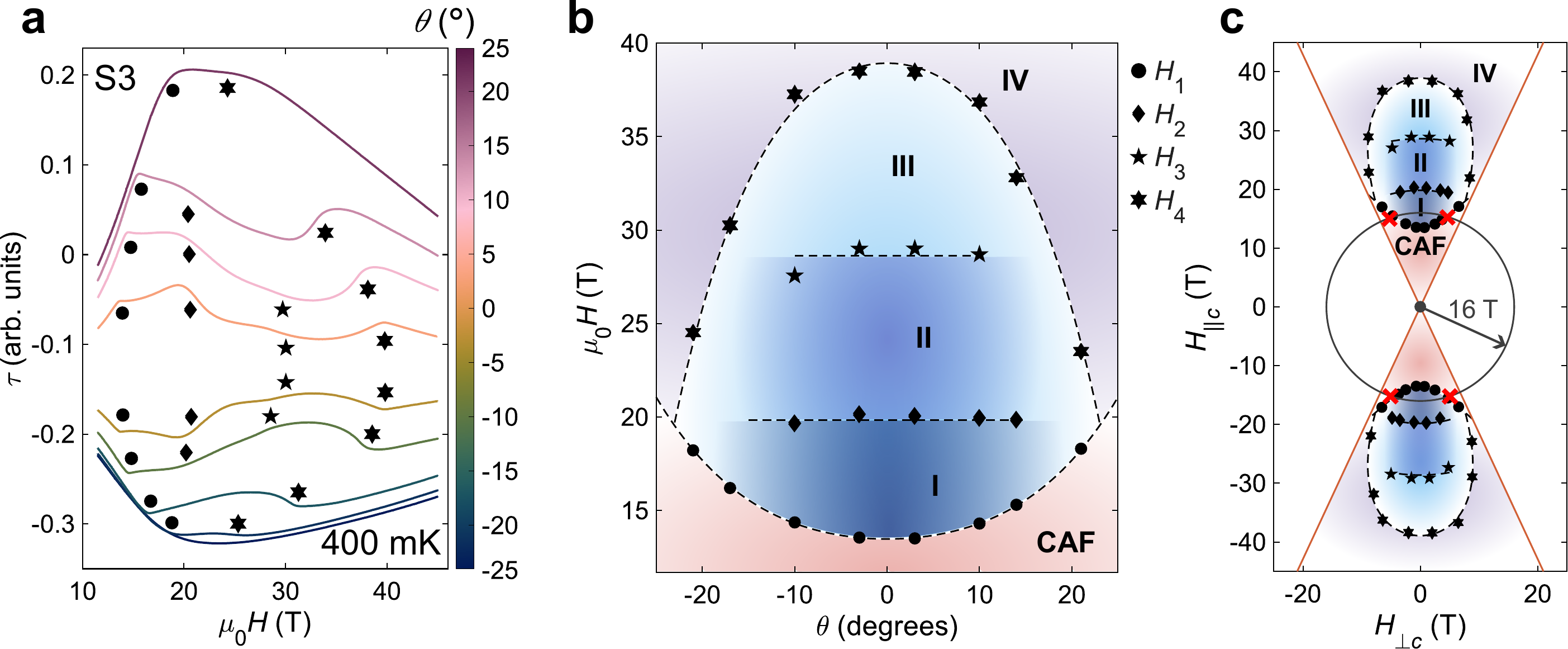}
\vspace{-0.2cm}
		\caption{
  {\bf Magnetic field dependence of the torque up to 45~T in different field orientations.}
  {\bf (a)} Field dependence of the magnetic torque at 400~mK measured at constant field angles, $\theta$, tilted away from the easy axis. The locations of four torque anomalies $H_{1-4}$ are indicated by circles, diamonds, stars and hexagrams, respectively.
  {\bf (b)} Angular dependence of the transition fields $H_{1-4}$. Dashed black lines are guides to the eye, dividing the experimentally observed CAF, I, II, III, and IV phases.
  {\bf (c)} Polar phase diagram constructed at 400~mK after imposing a two-fold symmetry axis perpendicular to the easy axis. The vertical and horizontal axes correspond to the applied field components parallel and perpendicular to the easy axis, respectively. The solid orange lines represent the angular measurement range of panel (a). The grey circle represents an example rotation trajectory in a field of 16~T, in correspondence with Fig.~\ref{Fig3}(a). Red crosses mark the positions of anomalies, $\theta_1$, detected in the angular dependence of the torque at 2~K in 16~T, as shown in Figs.~\ref{Fig2}(b) and \ref{Fig3}(a).
  }
  \label{Fig4}
\end{figure*}

An alternative approach to track changes in magnetic anisotropy inside different magnetic phases of $2H$-AgNiO$_2$ is to investigate the angular dependence of torque at constant temperatures while increasing the magnetic field strength in steps, as shown in Figs.~\ref{Fig3}(a-c).
In the paramagnetic phase at 150~K (Fig.~\ref{Fig3}(c)), the torque expectedly follows $\tau\propto \sin{2\theta}$ throughout the entire field range, while in the CAF phase at 2~K (Fig.~\ref{Fig3}(a)), a sawtooth tilt grows with field up to 14~T. Above 14~T, kinks emerge in the angular dependence of the torque near the easy axis, labelled $\theta_1$ in Fig.~\ref{Fig3}(a) in correspondence with Fig.~\ref{Fig2}(b). A turning point in the field dependence of the sawtooth parameter, $\gamma$, at 2~K is also observed (see SM Fig.~S10 \cite{SM}).

Interestingly, very close to the PM--CAF phase boundary at 19~K, the angular dependence of the torque acquires an additional modulation above 8~T, as shown in Fig.~\ref{Fig3}(b). This behaviour can be described alternatively through a Fourier decomposition: $\tau(\theta)=\sum_{n=1}^{\infty}A_{2n}\sin{2n(\theta-\phi_{2n})}$. The evolution of the two-fold ($A_2$) and four-fold ($A_4$) components is shown in Fig.~\ref{Fig3}(f). Above 8~T, the four-fold component, $A_4$, grows and is followed by a turning point in the two-fold component at 11~T.
Moreover, in the field range dominated by the four-fold component, the angular derivative $\partial\tau/\partial\theta$ is negative both along the easy axis and the honeycomb plane, indicating two orthogonal stable equilibrium field orientations.
This complex angular behaviour
is captured well by
simulations considering
magnetic transitions between 
phases with opposite susceptibility anisotropy, as will be  discussed later in Sec.~\ref{sec:simulations}.

\subsection{Angular-dependent phase diagram up to 45~T}

Next, we present
signatures in torque induced by varying magnetic fields in fixed sample orientations.
Fig.~\ref{Fig4}(a) shows the field dependence of the magnetic torque up to 45~T at 400~mK for different fixed field angles, $\theta$.
We observe four distinct anomalies, consistent with previous studies for measurements close to $H||c$ \cite{Coldea2014, Coldea2009}.
For $\theta\approxeq0^\circ$, the first anomaly at $H_1\approx14$~T is characterised by a change in slope of the torque and a peak in the second derivative, as shown in Fig.~S13 in the SM \cite{SM}.
The next two anomalies manifest as local maxima in the torque at $H_2\approx20$~T followed by a subtle kink at $H_3\approx28$~T.
These anomalies are characterised by peaks in the second and first derivatives of the torque, respectively (see Fig.~S13 in the SM \cite{SM}).
Finally, the fourth anomaly at $H_4\approx40$~T is marked by a change in slope of the torque and a peak in the first derivative (Fig.~S13 in the SM \cite{SM}).

As the magnetic field is tilted away from the easy axis and $|\theta|$ increases, the position of each anomaly is tracked to obtain its angular dependence, as shown in Fig.~\ref{Fig4}(b).
While $H_1$ and $H_4$ have strong angular dependences, the middle anomalies, $H_2$ and $H_3$, display hardly any shift in position with changing field orientation.
In addition, the signatures associated with $H_2$ and $H_3$ die away roughly $10^\circ$ from the easy axis, as seen in Fig.~\ref{Fig4}(a).
Meanwhile, the angular dependences of $H_1$ and $H_4$ are captured by quartic expressions of the form $H_i(\theta)=a_i(\theta^2+b_i)^2+c_i$, intersecting for $|\theta|\approx23^\circ$, as apparent in Fig.~\ref{Fig4}(b).

Based on these experimental data, we propose a polar phase diagram for $2H$-AgNiO$_2$ at $T=400$~mK in the range $|\theta|<25^\circ$ (assuming two-fold rotational symmetry), shown in Fig.~\ref{Fig4}(c). Within this angular range, the first and fourth anomalies bound an egg-shaped region that contains the proposed phases I, II, and III. Above $H_4$, no other anomalies have been reported even up to 90~T along the easy axis \cite{Coldea2014}.

\begin{figure*}[hbtp]
	\centering
\includegraphics[width=0.9\linewidth, clip=true]{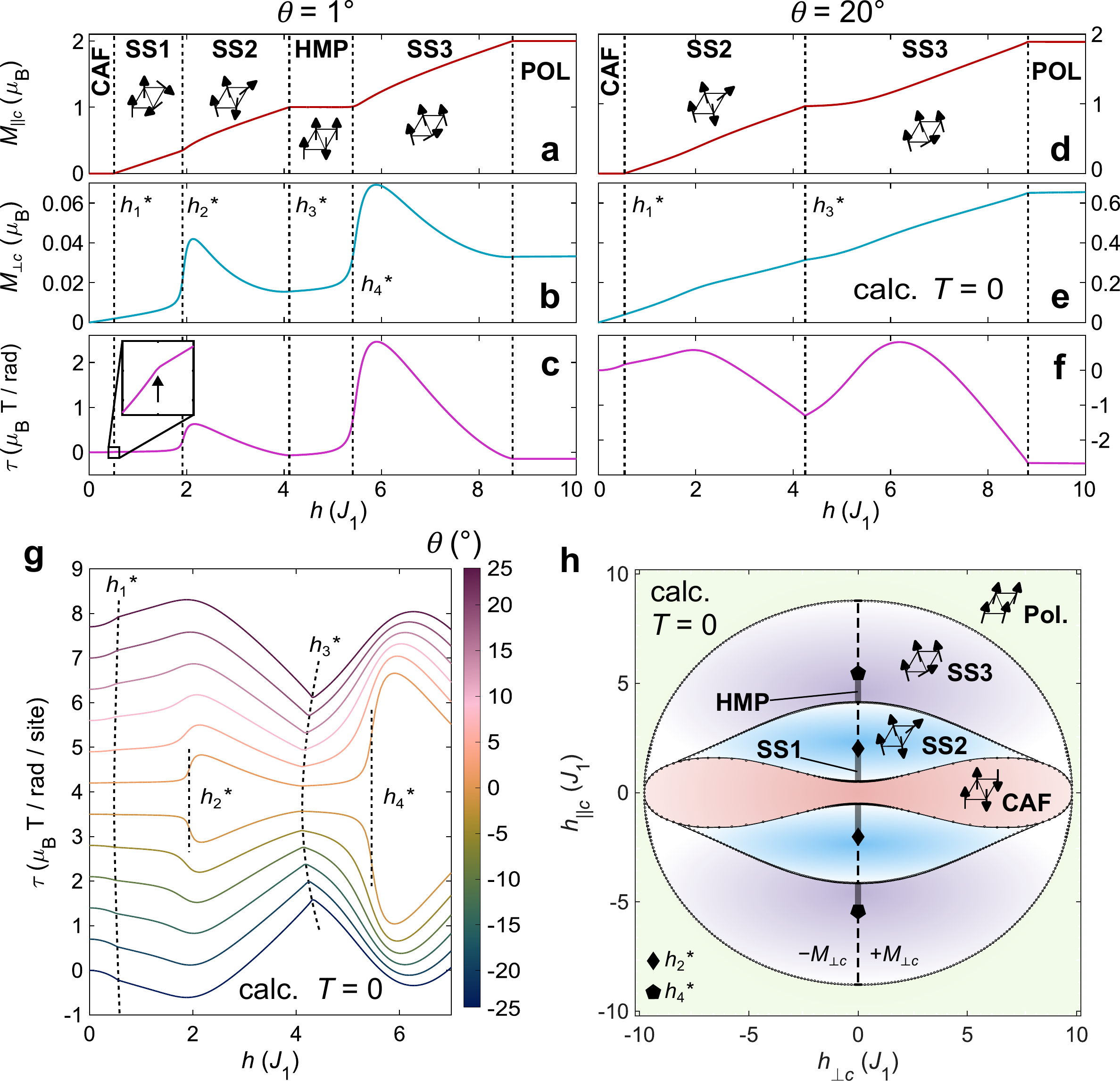}
\vspace{-0.2cm}
		\caption{
  {\bf Polar phase diagram at absolute zero for the $J_1$-$J_2$-$D$-$J_\perp$ model.}
    {\bf (a-c)} Mean field calculations of the magnetisation components (in units $\mu_B$ per spin) parallel and perpendicular to the easy axis, and the magnetic torque, respectively, for constant $\theta=1^\circ$ away from the easy $c$ axis. The inset in panel (c) shows a zoom-in of the torque at the field $h_1^*$, with an arrow to highlight the transition.
    {\bf (d-f)} Calculations analogous to panels (a-c) performed for constant $\theta =20^\circ$ away from the easy axis.
    {\bf (g)} Field dependence of the magnetic torque for different angles, $\theta$, away from the easy axis. The transition fields $h_1^*$ and $h_3^*$ represent the CAF--SS2 and SS2--SS3 boundaries, respectively. $h_2^*$ and $h_4^*$ are the SS1--SS2 and HMP--SS3 transitions and are only seen very close to $\theta=0$.
    {\bf (h)} Mean-field polar phase diagram. The vertical and horizontal axes represent the applied field components parallel and perpendicular to the easy axis, respectively, in units of the strongest exchange $J_1$. The solid black lines represent phase boundaries deduced in accordance with the calculations in panels (a-g). The solid diamonds and pentagons represent the transition fields $h_2^*$ and $h_4^*$, respectively. The SS1 and HMP phases along the easy axis direction are highlighted with shaded vertical strips within the SS2 and SS3 regions. The vertical dashed lines inside the SS2 and SS3 supersolid regions are boundaries which discontinuously change the sign of the planar magnetisation $M_{\perp c}$ when crossed.
    }
  \label{Fig5}
\end{figure*}

\begin{figure*}[hbtp]
	\centering
\includegraphics[width=0.9\linewidth, clip=true]{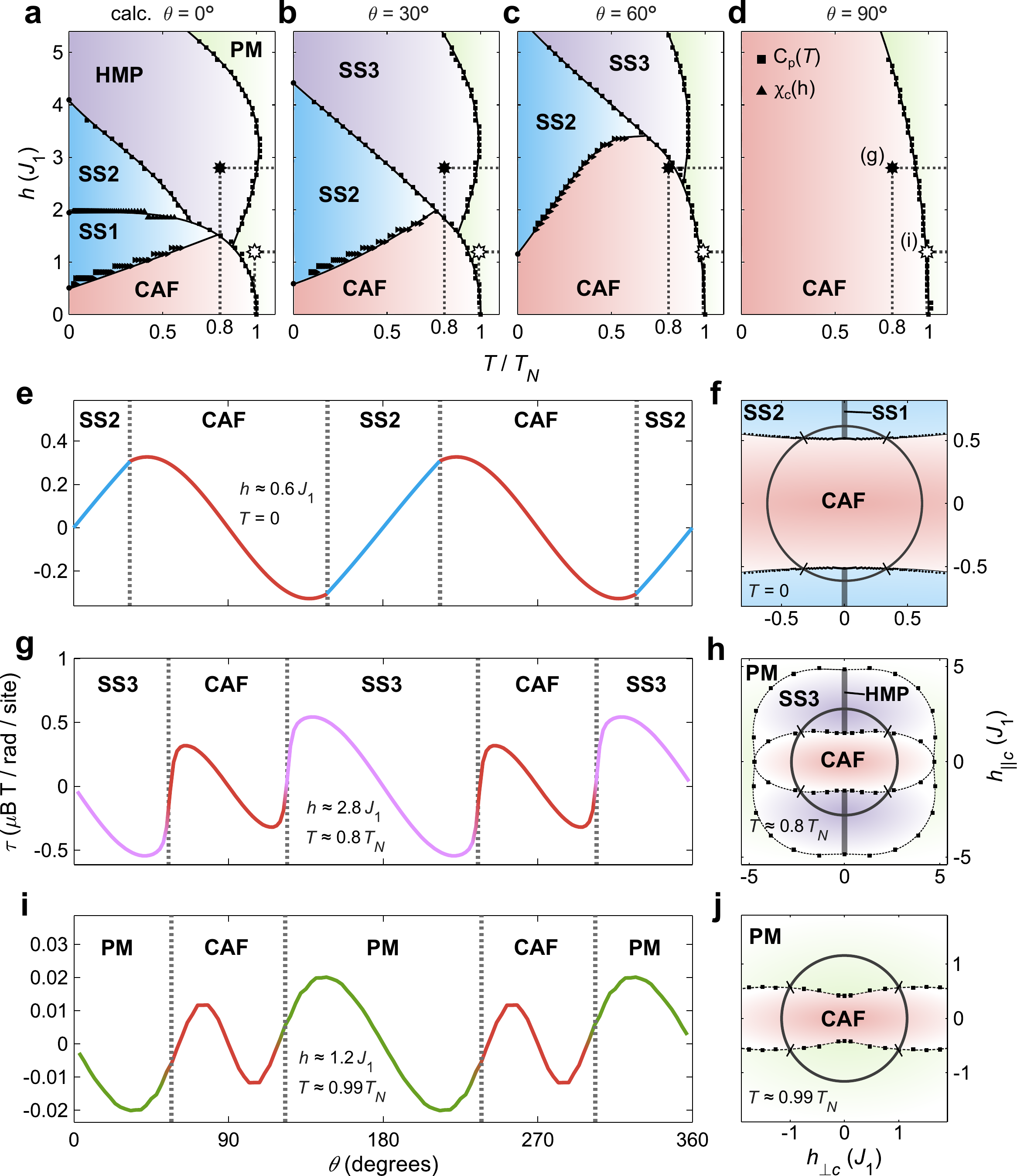}
\vspace{-0.2cm}
		\caption{
  {\bf Monte Carlo simulations of the $J_1$-$J_2$-$D$-$J_\perp$ model and comparison with experiments on $2H$-AgNiO$_2$.}
  {\bf (a-d)} $h$--$T$ phase diagrams for different field orientations simulated using Monte Carlo methods (see also Refs.~\cite{Seabra2010,Seabra2011,Coldea2014}). $\theta$ is the angle subtended by the applied field and the easy axis. Solid squares and triangles represent anomalies in the heat capacity and magnetic susceptibility, respectively.
  Solid black boundaries are guides to the eye. The black and white stars in each phase diagram correspond to the field and temperature values of the angular sweeps shown in panels (g) and (i), respectively.
  {\bf (e, g, i)} Simulated angular dependence of the torque at $T=0$, 0.8~$T_\text{N}$, and 0.99~$T_\text{N}$, respectively. Segments are coloured red, blue, purple, or green to signify where the system is in the CAF, SS2, SS3, or PM phases, respectively. The vertical dashed grey lines represent locations of phase transitions induced by field rotation.
  {\bf (f, h, j)} Phase diagrams corresponding to panels (e), (g), and (i), respectively. Solid circles represent the field rotation trajectories of the angular sweeps shown in (e, g, i). Shaded strips, labelled SS1 and HMP in panels (f) and (h), represent the SS1 and HMP phases which exist only along $h\parallel c$.
    }
  \label{Fig6}
\end{figure*}

\subsection{Classical spin simulations and phase diagram}
\label{sec:simulations}
To understand our torque data, we employ semi-classical calculations on a localised antiferromagnetic $J_1$-$J_2$ triangular Heisenberg model with easy axis single-ion anisotropy \cite{Seabra2011, Coldea2014, Wheeler2009} (Hamiltonian available in the SM \cite{SM}).
We consider the following exchange parameters: $J_1, J_2=1.00, 0.15$~meV as the first and second-nearest intralayer neighbour couplings, 
$J_\perp=-0.15$~meV as the ferromagnetic interlayer coupling, and single-axis anisotropy $D=0.25$~meV. These parameters were chosen to be consistent with previous Monte Carlo studies \cite{Coldea2014}. The g-tensor is assumed isotropic with $g=2$, and the applied magnetic field is $\mathbf{h}=g\mu_B\mu_0\mathbf{H}$, with magnitude denoted by $h=|\mathbf{h}|$.

The easy-axis anisotropy $D$ is a crucial tuning parameter in stabilising supersolid phases \cite{Seabra2011}. 
Here, we have chosen $D/J_1=0.25$ to reflect the anomalies observed in our torque data (Fig.~\ref{Fig4}(a)) and to maintain consistency with previous studies \cite{Coldea2014, Wawrzynska2008}. The role of $D$ is to pin the orientation of the spins to the crystallographic $c$ axis, where $D\rightarrow\infty$ brings the system towards the Ising limit \cite{Seabra2011}.
As previous studies \cite{Seabra2010, Seabra2011, Seabra2011_2} have shown, for certain values of $D$, additional first-order transitions can be realised in the phase diagram, e.g. to a traditional spin-flop state or a one-third magnetisation plateau \cite{Seabra2010, Seabra2011_2}.
Additional phase diagrams, similar to Fig.~\ref{Fig4}(c), for varying degrees of single-ion anisotropy can be found in Fig.~S2 in the SM \cite{SM}.

{\bf Mean-field calculations at absolute zero.}
Firstly, we focus on the theoretical signatures of the field-induced transitions for fields applied close to the easy axis, $h||c$ ($\theta \lesssim 1^{\circ}$). Figs.~\ref{Fig5}(a) and (b) capture the field dependence of the magnetisation components, $M_{\parallel c}$ and $M_{\perp c}$,
parallel and perpendicular to the easy $c$ axis, respectively, whereas
the resultant torque is shown in Fig.~\ref{Fig5}(c).
The field dependence of the calculated magnetisation and torque reveals four second-order field-induced transitions, $h^*_{1-4}$, up to the fully polarised state, as illustrated in Figs.~\ref{Fig5}(a-c) (dashed vertical lines).
The first transition at $h_1^*$ is from the CAF phase to the supersolid SS1 phase (see Fig.~\ref{Fig1}(c)). Here, an increase is observed in $M_{\parallel c}$, where the down spins of the CAF phase begin to cant in alternating directions to form the SS1 structure.
\cite{Seabra2010, Seabra2011}. 
Due to the slight tilt between the easy axis and the applied field ($\theta = 1^{\circ}$), a kink also appears in the magnetic torque at $H_1^*$, as depicted in the inset of Fig.~\ref{Fig5}(c).
Next, the system undergoes a transition between two magnetic supersolid phases, SS1 and SS2, at $h_2^*$ \cite{Seabra2010,Seabra2011}.
At the transition, a slight enhancement in
$M_{\parallel c}$ is observed, whereas $M_{\perp c}$ displays a peak in the first derivative. Just above $h_2^*$, a broad maximum is observed in $M_{\perp c}$ that propagates through to the magnetic torque, as evident in Figs.~\ref{Fig5}(b) and (c).
Semi-classically, this signature arises as half of the canted down spins in the SS1 phase rotate upwards towards the triangular plane upon entering the SS2 phase. The remaining down spins instead cant back towards the negative easy $c$ axis, giving rise to the 2:1:1 canted structure illustrated in Fig.~\ref{Fig1}(c).
The third transition field, $h_3^*$, represents the transition to the half-magnetisation plateau (HMP), accompanied by sharp changes in the slopes of $M_{\parallel c}$, $M_{\perp c}$ and the resulting torque,  
as shown in Figs.~\ref{Fig5}(a-c). Finally, crossing the fourth transition at $h_4^*$, the system changes from the HMP structure to the 3:1 canted supersolid, SS3, illustrated in Fig.~\ref{Fig1}(c).
Here, there is an increase in $M_{\parallel c}$ and a peak in the derivative of $M_{\perp c}$, as evident in Figs.~\ref{Fig5}(a) and (b). Slightly above $h_4^*$, a broad maximum is seen in both $M_{\perp c}$ and the torque, analogous to that near the SS1--SS2 ($h_2^*$) transition, as the down spins in the HMP phase rotate upwards towards the triangular plane upon entering the SS3 phase.
Inside the SS3 phase, further increasing the field gradually relaxes the spins into the fully polarised state, where all spins become parallel.

Note that a finite spontaneous magnetisation component perpendicular to the easy axis, $M_{\perp c}$, in the SS2 and SS3 phases persists even for magnetic fields aligned exactly parallel to the easy axis \cite{Seabra2010,Seabra2011}. An infinitesimal field tilt selects the direction of this spontaneous transverse moment in the $ab$ plane. Consequently, the sign of $M_{\perp  c}$ (and torque) will flip discontinuously if the field is rotated through $\theta=0^\circ$ above $h_2^*$ or $h_4^*$ within the SS2 or SS3 phases, respectively (see Fig.~S7 in the SM \cite{SM}). These transitions are represented in the polar phase diagram in Fig.~\ref{Fig5}(h) by vertical dashed boundaries along $h||c$ within the SS2 and SS3 regions. 

Next, we discuss
the theoretical signatures of different transitions when the magnetic field is tilted significantly away from the easy $c$ axis ($\theta=20^\circ$).
The field dependence of the different magnetisation components
and the resultant torque is displayed in Figs.~\ref{Fig5}(d-f).  
The system now undergoes two field-induced transitions, before reaching the field-polarised state,
as opposed to the four observed for field aligned close to the easy axis (Figs.~\ref{Fig5}(a-c)). 
In this orientation, only $h_1^*$ and $h_3^*$ survive, appearing as clear kinks in both $M_{\parallel c}$ and the calculated torque.
This is because the SS1 and HMP phases are only truly stabilised for $h||c$, and as special cases of the more general 2:1:1 (SS2) and 3:1 (SS3) canted spin structures, respectively.
As a consequence, the SS1--SS2 ($h_2^*$) and HMP--SS3 ($h_4^*$) transitions become ill-defined when tilting the magnetic field away from the easy $c$ axis, as shown in Fig.~\ref{Fig5}(g).

{\bf Monte Carlo simulations at finite temperature.}
In order to understand the 
evolution of supersolid phases as a function of temperature, we implement a parallel tempering Monte Carlo simulation on the minimal easy-axis model, after Ref.~\cite{Seabra2011}.
The phase boundaries are characterised by peaks in the simulated heat capacity 
 and anomalies in the magnetic susceptibility (see Fig.~S4 in the SM \cite{SM}). However, these transitions can alternatively be characterised by a suitably defined $U(1)$ symmetry-breaking order parameter, staggered magnetisation, or spin stiffness, as detailed in Ref.~\cite{Seabra2011}.

The resulting field-temperature ($h$--$T$) phase diagram
for $h||c$ ($\theta=0^\circ$) is calculated up to $h=5.3J_1$, as shown in Fig.~\ref{Fig6}(a).
 In this field range, we detect the expected
 four magnetic phases which have
 a temperature dependence consistent with previous Monte Carlo studies \cite{Coldea2014, Seabra2011}.
As the field is tilted away from the easy $c$ axis ($\theta\geq15^\circ$),
the phase boundaries of the canted supersolid phases move towards
higher magnetic fields, and phases such as SS1 and HMP are 
not stabilised any more (see Figs.~\ref{Fig6}(b-c)).
  Finally, when magnetic field is aligned exactly perpendicular to the easy axis ($\theta=90^\circ$), only the CAF phase is present, as evident in Fig.~\ref{Fig6}(d). Additional  $h$--$T$ phase diagrams for 
other angles between $\theta = 0^\circ$ and $90^\circ$ are shown in Fig.~S5 in the SM \cite{SM}.

\subsection{Comparison of torque experiments with simulations}

Having detailed experimental data and theoretical calculations, we can attempt to develop a qualitative understanding of the angular dependence of the experimentally observed torque.
The theoretical polar phase diagram shown in Fig.~\ref{Fig5}(h) contains a multitude of phases with a peanut-shaped CAF region surrounded by elliptical canted supersolid regions.
As discussed in Sec.~\ref{sec:simulations}, two of the phase transitions, SS1--SS2 at $h_2^*$ and HMP--SS3 at $h_4^*$, are present only close to the easy $c$ axis, $h||c$, and are suppressed by tilting the magnetic field.
Comparing with the experimental polar phase diagram in Fig.~\ref{Fig4}(c),
we observe that the boundary traced by the anomaly $H_1$ resembles the concave curvature of the CAF--SS2 boundary, $h_1^*$, close to the easy $c$ axis. Likewise, the measured $H_4$ boundary possesses a curvature with the same sign to that of the convex elliptical SS2--SS3 boundary, $h_3^*$. Meanwhile, the boundaries traced by the intermediate anomalies, $H_2$ and $H_3$, span only a narrow angular window ($|\theta|\lesssim10^\circ$), within which the radial positions of the anomalies hardly vary (see Fig.~\ref{Fig4}(b)).
We note that the limited angular window over which $H_2$ and $H_3$ are present echoes the behaviour of the $h_2^*$ and $h_4^*$ transitions, which likewise occur only for orientations very close to $h||c$.

To further assess the effect of crossing different phase boundaries, we consider the field dependences of the measured and calculated torque over a range of fixed field orientations, shown in Figs.~\ref{Fig4}(a) and \ref{Fig5}(g), respectively.
The simulated CAF--SS1 transition manifests as a kink in the torque at $h^*_1$, resembling the experimental anomaly at $H_1$, which is itself characterised by a change in slope and a peak in the second derivative (Fig.~S13 in the SM \cite{SM}). 
Another qualitative comparison can be drawn between the local maximum observed in the torque in the vicinity of $H_2$, close to $\theta=0$, and the broad maximum located just above the transition $h_2^*$, as shown in Fig.~\ref{Fig5}(c).
Our comparison of torque signatures can be extended to the fourth anomaly at $H_4$ and the SS2--SS3 transition at $h_3^*$, which both manifest qualitatively as changes in slope.
However, the anomaly $H_3$, unlike $H_{1,2,4}$, does not have a qualitative counterpart within the current minimal model, even over other single-ion anisotropy values considered (see Fig.~S2 in the SM \cite{SM}). This suggests that $H_3$ may reflect physics beyond the present model, such as the itinerant effects proposed in Ref.~\cite{Coldea2014}, or from additional terms absent from the present spin Hamiltonian \cite{Wheeler2009}. Furthermore, the present semi-classical treatment does not incorporate the quadrupolar moments of the $S=1$ spins, which may require a more realistic description based on a $u(3)$ spin representation \cite{Remund2022}.

In addition to the magnetic torque, the previously reported magnetisation of $2H$-AgNiO$_2$ for $H||c$ \cite{Coldea2014} reveals clear signatures related to the first two torque anomalies, $H_1$ and $H_2$ (Fig.~\ref{Fig4}(a)). We note that the reported increase in magnetisation at $H_1$ \cite{Coldea2014} echoes the behaviour of the calculated magnetisation $M_{\parallel c}$ at the CAF--SS1 transition ($h_1^*$), shown in Fig.~\ref{Fig5}(a). Moreover, the magnetisation at $H_2$ appears to undergo a slight enhancement \cite{Coldea2014}, which may be qualitatively compared with the enhancement of $M_{|| c}$ across the SS1--SS2 transition ($h_2^*$).

Next, we assess the origin of different signatures observed in the angular dependence of the torque (Figs.~\ref{Fig3}(a) and (b)) based on our simulations at different temperatures.
Mean-field calculations at $T=0$ suggest that the anomalies observed in the experimental high-field angular dependence ($\theta_1$ in Fig.~\ref{Fig2}(b)) resemble the signatures of phase transitions caused by rotating through the CAF--SS2 phase boundary, as shown in Figs.~\ref{Fig6}(e) and (f).
These rotation-induced phase transitions are characterised by discontinuities in the slope of the torque which appear close to boundary crossings. The triangular-wave shape of the angle dependence arises due to the torque changing from a quasi-linear shape in the canted supersolid SS2 phase to a more sinusoidal shape in the CAF phase, as illustrated in Fig.~\ref{Fig6}(e).
However, the model does not fully reflect the degree of sawtooth tilt ($\gamma$ in Eqn.~\ref{eqn:sawtooth}) seen experimentally in fields below the CAF--I boundary (see Fig.~\ref{Fig3}(a) and SM Fig.~S10 \cite{SM}).

Sawtooth behaviour in torque can be controlled by tuning the strength of anisotropic spin Hamiltonian terms \cite{Riedl2019}
and has been observed in various other magnetic systems, even in the absence of phase transitions \cite{Ma2024, Pearce2024}. 
The dependence of the sawtooth tilting parameter, $\gamma$ (from Eqn.~\ref{eqn:sawtooth}), on the single-ion anisotropy $D$ is explored with mean-field calculations in Fig.~S3 in the SM \cite{SM}.
Interestingly, within the CAF phase, we identify an empirical scaling relation, $\gamma-1=\alpha h^2/(DJ_1)$ with $\alpha\approx1/10$. Hence, at fixed field, enhancing the single-ion anisotropy $D$ reduces the sawtooth tilt. At the same time, increasing $D$ expands the CAF phase (see Fig.~S2 in the SM \cite{SM}): linear spin-wave theory gives $h_1^*(0)=2D$ for the CAF--SS1 transition along $h|| c$, driven by the closure of a magnon gap at a soft point \cite{Wheeler2009}. Therefore, while a larger $D$ may reduce $\gamma$ at fixed $h$, it permits higher fields, and hence stronger sawtooth tilts, to be attained before the supersolid phase is entered.

Monte Carlo simulations at finite temperatures 
indicate how additional modulations can occur in the angular dependence of torque, resembling those observed in our measurements close to $T_\text{N}$ (see Fig.~\ref{Fig3}(b)).
The magnetic phase diagrams at $T=0.8~T_N$ and $0.99~T_N$ show that these extra modulations come from
transitions between the CAF phase and either the SS3 or PM phases under field rotation, respectively, as shown in Figs.~\ref{Fig6}(g-j). Since the susceptibility anisotropy $(\chi_{a}-\chi_c)$ in the SS3 and PM phases has opposite sign to that in the CAF phase, this causes free energy minima, $\partial\tau/\partial\theta<0$, to appear for field both parallel and perpendicular to the easy axis. Experimentally, a narrow region of an ordered magnetic phase (see IV in Fig.~\ref{Fig1}(d)) appears between the PM and CAF phases in high fields along $H||c$, as established by previous heat capacity studies \cite{Coldea2014, Coldea2009}. Previous studies suggest that this phase could indeed have characteristics resembling the HMP or SS3 phases \cite{Coldea2014}, although the torque angular dependence may equally be understood through a similar phase switching mechanism between the CAF and PM phases. The calculated angular dependence of the torque in different constant fields is shown in Fig.~\ref{Fig3}(e) for $T=0.8~T_N$, and Fig.~S6 for $T=0.99~T_N$ in the SM \cite{SM}.

Note that, in previous studies along $H||c$, the anomalies $H_{2}$ and $H_3$ were both attributed to the influence of the itinerant Ni$^{3.5+}$ sublattice electron spins \cite{Coldea2014}. Meanwhile, $H_1$ and $H_4$ were interpreted as the CAF--SS1 ($h_1^*$) and SS1--SS2 ($h_2^*$) transitions, respectively. These assignments were based on a detailed comparison between the temperature dependences of the experimental and theoretical phase boundaries for $H||c$, scaled relative to a tricritical point \cite{Coldea2014} (see Fig.~\ref{Fig1}(d)). Here, we consider the magnetic field angle as another tuning parameter to assess how the experimental anomalies evolve with orientation, providing a basis and supplementary constraint for alternative interpretations within the local moment model.

Magnetotransport measurements on a single crystal \cite{Coldea2014}
show field-induced signatures 
which complement the first three anomalies $H_{1-3}$.
There is strong evidence to suggest that the primary electronic scattering channel in $2H$-AgNiO$_2$ is through low-energy spin fluctuations, as evident in the suppression of resistivity below $T_\text{N}$ where a magnon gap opens \cite{Coldea2014, Wheeler2009}. 
Calculations have predicted that the magnetic moment of the metallic Ni$^{3.5+}$ sites is small but finite (of order $\sim0.1~\mu_B$) \cite{Coldea2014, Wawrzynska2008}, and therefore may influence the overall magnetic interactions in this system.
To verify to what extent the observed transitions are due to itinerant effects, as opposed to local moments, further in-depth investigations of the transport properties of $2H$-AgNiO$_2$ may be instructive.

\section{Conclusion}

We have studied the magnetic anisotropy of $2H$-AgNiO$_2$ using angle-dependent torque in magnetic fields up to 45~T and temperatures down to 400~mK in different crystallographic orientations. 
In the angular dependence of the torque, in the collinear antiferromagnetic phase, we found sawtooth behaviour and anomalies close to the easy $c$ axis in high fields. Additionally, close to $T_\text{N}$, we detected an unexpected additional modulation in the torque angular dependence. In the torque field dependence, we tracked four field-induced anomalies and constructed a polar phase diagram at 400~mK.

Our measurements were compared with simulations considering a $J_1$--$J_2$ model with moderate single-ion anisotropy. We performed mean-field calculations and finite-temperature Monte Carlo simulations to construct phase diagrams at different temperatures and in different magnetic field orientations. The resulting polar phase diagram at absolute zero 
contains four field-induced transitions close 
to the easy $c$ axis, whereas only two transitions
survive when the magnetic field is rotated away.
Monte Carlo simulations enabled us to 
build detailed polar phase diagrams at finite temperature.
We found that the anomalies observed in the angular dependence of the torque at low temperatures, in high fields, bear resemblance to simulated transitions induced by field rotation between the collinear antiferromagnetic and 2:1:1 canted supersolid phases. 
Similarly, the additional modulation in the angular dependence observed close to $T_\text{N}$ resembled
the rotation-induced transitions between the collinear antiferromagnet and either the 3:1 canted supersolid or paramagnetic phases at finite temperature.
In summary, our results demonstrate that the phase diagram of $2H$-AgNiO$_2$ is highly sensitive to field orientation, and that phase boundary crossings can give rise to rich signatures in the magnetic torque which are qualitatively captured by a local-moment description.

\section{Acknowledgments}

This work was partially supported by the EPSRC (EP/I004475/1) and UKRI3274,
the Oxford Centre for Applied Superconductivity
and the ISABEL project of the European Union’s Horizon's 2020 Research and Innovation Programme Grant Agreement Number No 871106. 
Part of this work was supported by HFML-RU and LNCMI-CNRS, members of the European Magnetic Field Laboratory (EMFL) and by EPSRC (UK) via its membership to the EMFL (grant no. EP/N01085X/1).
A portion of this work was performed at the National High Magnetic Field Laboratory, which is supported by National Science Foundation Cooperative Agreement No. DMR-1644779 and DMR-2128556 and the State of Florida.
We acknowledge financial support from the Oxford University John Fell Fund.
RC acknowledges support from the European Research Council under the European Union’s Horizon's 2020 Research and Innovation Programme Grant Agreement Number 788814 (EQFT). 
J.S.P  acknowledges financial support from the EPSRC studentship EP/W524311/1, scholarship funding from the Department of Physics, via OxPEG, and the Leathersellers' Scholarship from St. Catherine's College, Oxford.
AIC acknowledges an EPSRC Open Fellowship (UKRI3274).

\section{Data Availability}
The data that support the findings of this study will be made available through the open access data archive at the University of Oxford (ORA) \cite{pearce2026data}.

\bibliography{AgNiO2_bib}

\end{document}